\documentclass[twocolumn,showpacs,preprintnumbers,amsmath,amssymb,superscriptaddress]{revtex4-1}

\usepackage{graphicx}
\usepackage{dcolumn}
\usepackage{bm}
\usepackage{dsfont}
\usepackage{color}  
\usepackage{dcolumn}      
\usepackage{array}
\usepackage{xcolor}

\begin{document}

\title{Mesoscopic thermoelectric transport near zero transmission energies
}
\author{A. Abbout}
\affiliation{Laboratoire CRISMAT, UMR 6508 CNRS, ENSICAEN et Universit\'e de Caen Basse Normandie, 6 Boulevard Mar\'echal Juin, F-14050 Caen, France}
\author{H. Ouerdane}
\affiliation{Laboratoire CRISMAT, UMR 6508 CNRS, ENSICAEN et Universit\'e de Caen Basse Normandie, 6 Boulevard Mar\'echal Juin, F-14050 Caen, France}
\affiliation{Universit\'e Paris Diderot, Sorbonne Paris Cit\'e, Institut des Energies de Demain (IED) URD 0001, 75205 Paris, France}
\author{C. Goupil}
\affiliation{Laboratoire CRISMAT, UMR 6508 CNRS, ENSICAEN et Universit\'e de Caen Basse Normandie, 6 Boulevard Mar\'echal Juin, F-14050 Caen, France}
\affiliation{Universit\'e Paris Diderot, Sorbonne Paris Cit\'e, Institut des Energies de Demain (IED) URD 0001, 75205 Paris, France}

\begin{abstract}
We study the thermoelectric transport coefficients of a one-dimensional (1D) electron waveguide connected to one and then two off-channel cavities, in the presence of dephasing phonons. The model system is that of a linear chain described as a 1D lattice. For simplicity we consider single-mode cavities, which may be tuned with external gates. While the presence of only one off-channel cavity yields a nearly symmetric transmission profile, which is canceled around the cavity mode, an additional cavity modifies this profile strongly and results in an asymmetric shape characterized by oscillations. In both cases, we consider electron-phonon interactions in our calculations and analyze their effects on the transmission function around the Fermi energy. Knowledge of the energy-dependent transmission function allows the numerical computation of thermoelectric transport coefficients, including the thermopower. In the presence of a second off-channel cavity, the sign of the thermopower depends on the relative position of this cavity energy level with respect to the Fermi energy: the thermopower is positive when low-energy electrons in the vicinity of the Fermi level are not transmitted, and becomes negative when the higher-energy electrons are not transmitted.
\end{abstract}

\pacs{72.10.-d, 72.15.Jf, 73.23.-b}
\keywords{electronic transport, scattering mechanisms, mesoscopic systems, thermoelectricity}

\maketitle

\section{Introduction}
Thermoelectric systems are heat engines which, by using the electron gas formed by their conduction electrons as a working fluid, directly convert a heat flux into electrical power and vice-versa, depending on the desired operation mode. In the linear regime, the coupling between heat and electricity is embodied in a parameter called thermopower or Seebeck coefficient, $S$. The principal virtues of thermoelectric heat engines are that their conversion efficiency is not size-dependent and that their operation does not rely on moving parts; the main drawback is their modest energy conversion efficiency, which is at best of the order of 10 \% of the efficiency of the ideal Carnot thermodynamic cycle.

Since the first works reported by Seebeck \cite{Seebeck}, Oersted \cite{Oersted}, and Peltier \cite{Peltier}, general interest in thermoelectricity has much varied with time. Some significant milestones in this field are Thomson's thermodynamic analysis \cite{Thomson}, Callen's \cite{Callen} adaptation of Onsager's formalism \cite{Onsager}, and Ioffe's works on device operation \cite{Ioffe}. In particular Ioffe related the thermoelectric conversion efficiency to a dimensionless parameter denoted $ZT$, where $T$ is the average temperature across the device and $Z$ is a parameter that depends on the material's properties, namely the thermopower $S$, and the electrical and thermal conductivities, $\sigma$ and $\kappa$ respectively: $ZT=\sigma S^2 T/\kappa$ ($\kappa$ entails both electron and lattice thermal conductivities, $\kappa_{\rm e}$ and $\kappa_{\rm lat}$, but we neglect the latter throughout the present work since we are mainly interested in the electronic transport properties \cite{remark1}). Though $ZT$ has a profound meaning on the thermodynamic level, this parameter is much used as a means to measure how well in terms of energy conversion efficiency, a device operates. A good performance is associated to high values of $ZT$ (typically around 1).

Enhancement of the performance can be obtained following three routes: $i/$ optimization of the thermoelectric material's properties \cite{handbook}; $ii/$ optimization of device working conditions through appropriate design \cite{handbook,Heikes1961} and electrical and thermal impedance matchings \cite{Apertet1}; and $iii/$ design of new nano- and mesostructures \cite{Dresselhaus,Shakouri,Trocha}. The first route, which essentially consists of finding means to decrease $\kappa$ and increase $\sigma$ has produced a number of very interesting results, but further significant progress seems out of reach \cite{Vinning}, one of the difficulties being that both $\sigma$ and the electronic part $\kappa_{\rm e}$ of the thermal conductivity $\kappa$ are tightly linked; this link is the Wiedemann-Franz law for metals \cite{Jonson}. The second line of work is based on the observation that the highest performances are usually obtained in a very limited temperature range for a given material, so that devices must be designed in such a way that they may operate over a large temperature range, which is usually achieved by segmentation; in addition, when seeking maximum efficiency at maximum output power for a thermoelectric generator, one should ensure that the conditions for both thermal and electrical impedance matchings are fulfilled.

The third route, which we explore in this paper, was permitted thanks to the rapid and tremendous progress in the field of fabrication of nano- and mesoscopic artificial structures. The interest of going down to such scales stems from a number of effects that truly enrich the physics of thermoelectricity and pave the way to possibly better device performance; we can mention quantum confinement \cite{Dresselhaus,Balandin,Yuan} and the related modification of the carriers' density of states, mesoscopic fluctuations, breakdown of the Fermi liquid picture \cite{Kane,Wakeham,Benenti}, and limits of validity of standard thermodynamic approaches due to finite-size effects. Further, since for macroscopic systems an increase of thermopower impacts on the values of thermal and electrical conductivities and may not necessarily result in a significant increase of $ZT$, it is also interesting to see how at the mesoscopic level the quantum effects pertaining to electron confinement and quantization of the transport coefficients \cite{Wees,Schwab} may modify the interplay between $S$, $\kappa$, and $\sigma$. In this paper we see how to achieve high figures of merit with systems having large thermopower and low thermal conductivity at the same time.

We focuse on a one-dimensional (1D) system in the presence of off-channel cavities, which perturb the electrical conductivity $\sigma$ or, equivalently, the energy-dependent transmission function ${\mathcal T}$. This system may be fabricated from structures which contain two-dimensional electron gases. The main property of the mesoscopic system we study is the cancellation of the transmission function at a given energy we may choose and vary using an external gate; other characteristics are described in Section 2. Now, the main idea underlying the present work comes from the Cutler-Mott formula \cite{Mott}, which gives the Seebeck coefficient at low temperature:

\begin{equation}
\label{Mott}
S=-\frac{\pi^2}{3} \frac{k_{\rm B}^2 T}{e} \left(\frac{\partial \ln(\sigma)}{\partial E}\right)_{E=E_{\rm F}}
\end{equation}

\noindent where the derivative of the electrical conductivity is taken at the Fermi energy $E_{\rm F}$. Equation \eqref{Mott} which is valid for metallic systems such as that under consideration in this paper, shows that when the transmission vanishes, the Seebeck coefficient tends to infinity. In general, when the transmission vanishes, the derivative $\partial\sigma/\partial E$ vanishes too, but in virtue of l'Hopital's rule, we know that the limit does diverge. Therefore, if the conduction band contains a transmission-cancelling energy level, the coupling between heat and electrical fluxes is enhanced in the vicinity of this energy, which in turn may yield a significant increase of the figure of merit $ZT$.

As will become obvious with the definitions of the transport coefficients given further below, a symmetric transmission function would tend to cancel the contributions from higher and lower energy terms with respect to the Fermi level, and hinder transport, so we seek configurations that makes the transmission function profile asymmetric. Indeed, the transmission function should be asymmetric around the Fermi energy so that the hot-reservoir electrons may leave their high energy states and proceed against the applied source-drain bias through the waveguide to the cold reservoir's available states as noted in Ref.~\cite{Linke}. In this work, we investigate the effects of electron-phonon coupling on the transmission profile, and see whether this coupling impacts positively on the transport coefficients and figure of merit.

\begin{figure}
\begin{center}
 \includegraphics[scale=0.3]{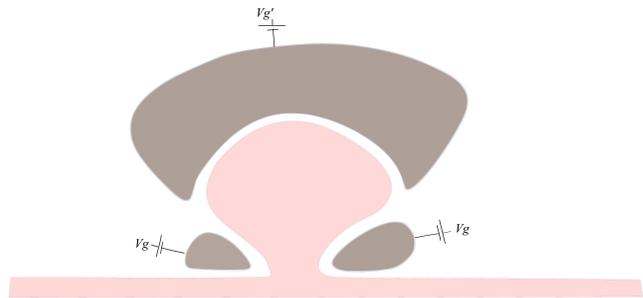}
\end{center}
\caption{(color online) An off-channel cavity connected to 1D electron waveguide. The shape of the cavity and its coupling to the waveguide are controlled by external gates.}
\label{Cavity}
\end{figure}

Our phenomenological model is based on a number of assumptions. First, our main framework is that of single-particle theory so that the ballistic electron transport in the waveguide may be described by a single-band effective mass Hamiltonian. We consider the linear response only, which implies that the temperature gradient across the system is small; incidently this assumption provides a small energy scale for the model. The temperature $T$ is sufficiently small so that the Cutler-Mott formula \eqref{Mott} remains valid. Effects of the couplings of the waveguide to the off-channel cavities and the phonons are studied with the nonequilibrium Green's function approach \cite{Datta}, adapting to our purposes recent works on dephasing effects on quantum transport \cite{Cresti,Datta2,Cresti2,Cresti3}. We assume that the system operates in a steady-state regime, and that the two reservoirs, to which the electron waveguide is connected, are perfect. The Fermi energy $E_{\rm F}$ in the whole system is given by the highest level occupied by an electron in either reservoirs at zero temperature. The phonons are assumed to be in equilibrium.

The article is organized as follows. In section II, we present in detail the basic system under consideration. The system Hamiltonian, the thermoelectric transport coefficients and the transmission function are defined and discussed considering the presence of only one off-channel cavity coupled to one site of the lattice. In section III, we introduce the effects of the coupling of parts of the 1D lattice to a bath of dephasing phonons; for this, we redefine our basic chain model using a decimation procedure. In section IV, we study the response of the system's thermoelectric transport coefficients to the presence of an additional off-channel cavity. The article ends on concluding remarks, followed by the Appendix where some detail on the calculation of the self-energies using the decimation procedure is given. 

\section{Model of the basic system}
We consider a 1D electron waveguide, which we describe as a 1D lattice; electron conduction is characterized by hopping between the lattice sites. The waveguide is connected first to one off-channel cavity by coupling to one of the lattice sites. We also assume a small cavity size so that it contains only one energy level accessible to the conduction electrons \cite{remark2} and, in the absence of the off-channel cavity, the system is uniform and has full transmission: ${\mathcal T}(E)=1, ~\forall E$, at any Fermi energy. The cavity is controlled by external gates which handle its coupling to the electron waveguide and its shape as shown in Fig.~\ref{Cavity}.

\subsection{Hamiltonian}
The Hamiltonian of the system can be written in a tight binding model as the sum of three contributions:

\begin{equation}
H=H_{\ell}+H_{\rm d}+H_{\rm c}
\end{equation}

\noindent where $H_{\ell}$ is the electron waveguide Hamiltonian, which describes the transport of an electron from $-\infty$ to $+\infty$ in the absence of the off-channel cavity:

\begin{equation}
H_{\ell}=\sum_{x=-\infty}^{\infty}  -t\left(|x\rangle\langle x+1|+|x+1\rangle\langle x|\right) +\epsilon_0 |0\rangle\langle0|
\end{equation}

\noindent where $|x\rangle$ is a Wannier state at location $x$ along the 1D lattice, and $t$ is the hopping coefficient, which we take as: $t = \hbar^2/2m^{\star}a^2$ in the effective mass approximation ($m^{\star}$ is the electron effective mass, and $a$ the lattice spacing). The energy $\epsilon_0$ is that of the site at the center of lattice ($x=0$). The dispersion relation derived from the Hamiltonian $H_{\ell}$ reads $E=-2t\cos(k)$, leading to a conduction band $E\in [-2t,+2t]$. The term $H_{\rm d}$ characterizes the off-channel cavity with a single level:

\begin{equation}
H_{\rm d} =V_{\rm d} |d\rangle\langle d|
\end{equation}

\noindent where the energy $V_{\rm d}$ of the state $|d\rangle$ can be controlled by an external gate. The term $H_{\rm c}$ characterizes the coupling between the cavity and the site at the center of the electron waveguide to which it is coupled:

\begin{equation}
H_{\rm c}=t_{\rm c}|0\rangle\langle d|+t_{\rm c}|d\rangle\langle 0|
\end{equation}

\noindent where the coupling parameter is noted $t_{\rm c}$.

\noindent This simple model may represent a multilevel quantum dot when the level spacing is large compared to the range of energies contributing to the transport due to temperature smearing. In all that follows, the hopping coefficient in the leads is taken as $t=1$ such that all the energies in our study are expressed in the units of $t$.

\subsection{Thermoelectric transport}
We consider three transport coefficients: the electrical conductivity $\sigma$, the thermopower $S$, and the electronic thermal conductivity $\kappa_0$, which is related to the zero-electric-current thermal conductivity $\kappa_{\rm e}$ \emph{via}:

\begin{equation}
\kappa_0=\kappa_{\rm e}+T\sigma S^2
\end{equation}

\noindent These three coefficients are given by the solutions of the Boltzmann equation \cite{Mahan,Kim}:

\begin{equation}\label{conductance}
\sigma= \int \left(-\frac{\partial f}{\partial E}\right) \mathcal{T}(E){\rm d}E
\end{equation}

\begin{equation}\label{Seebeck}
T\sigma S= \int(E-\mu)\left(-\frac{\partial f}{\partial E}\right) \mathcal{T}(E){\rm d}E  
\end{equation}

\begin{equation}\label{Kappa}
T\kappa_0= \int  (E-\mu)^2 \left(-\frac{\partial f}{\partial E}\right) \mathcal{T}(E){\rm d}E
\end{equation}

\noindent where $\mu$ is the electrochemical potential of the electron system. The conductivity $\sigma$ is expressed in units of $e^2/h$, with $h$ being the Planck constant, and $e$ being the electron charge. The important quantity that determines all the necessary physical quantities for our work is the transmission function $\mathcal{T}$. Knowledge of the three transport coefficients thus permits an analysis of the figure of merit $ZT$, which for a standard system where heat is transported by electrons and the lattice, may be rewritten as:

\begin{equation}\label{ZTe}
ZT=\frac{\displaystyle \sigma T S^2/\kappa_0}{\displaystyle 1-\sigma T S^2/\kappa_0} \times \frac{1}{1+\kappa_{\rm lat}/\kappa_{\rm e}}
\end{equation}

\noindent Note that in the present work, we focuse on the electronic thermal conductivity $\kappa_0$ because the 1D system is made from 2D electronic gases. So the figure of merit we compute and analyze in this article is the electronic part of $ZT$ given in Eq.~\eqref{ZTe}, neglecting the contribution of $\kappa_{\rm lat}$. For ease of notations, we retain $ZT$ as the electronic part of the figure of merit.

Equations \eqref{conductance}, \eqref{Seebeck}, and \eqref{Kappa} show that the electron waveguide, \emph{via} its transmission function ${\mathcal T}$ which we define below, acts as an energy-selective filter \cite{Linke}. As mentionned in the Introduction, for electron transport to take place, it is necessary that the transmission function be asymmetric, and this is clear with Eq. (\ref{Seebeck}) : since $(E-\mu)$ changes sign around the Fermi energy, a symmetric transmission function would yield cancellation. Also note that $(E-\mu)$ corresponds to the heat involved in the transport process, so a high energy conversion efficiency may be obtained on the conditions that the transmission profile permits a high eletrical flux from hot to cold reservoirs with a minimal heat flux.
 
\subsection{Transmission function}
The energy-dependent transmission function may be obtained from the Fisher-Lee formula \cite{FisherLee}:

\begin{eqnarray}\label{Landauer}
\mathcal{T}(E)=\Gamma_{\rm L} G \Gamma_{\rm R} G^\dagger   
\end{eqnarray}

\noindent where $G$ is the retarded single-particle Green's function of the central site (Wannier state $|0\rangle$ with energy $\epsilon_0$), taking into account the effect of the leads and the off-channel cavity by means of their self energies:

\begin{equation}\label{GF}
G=\frac{1}{E-\epsilon_0-\Sigma_{\rm L}-\Sigma_{\rm R}-\frac{\displaystyle t_{\rm c}^2}{\displaystyle E-V_{\rm d}}}
\end{equation}

\noindent where the self energies of the left ($\Sigma_{\rm L}$) and the right($\Sigma_{\rm R}$) leads are taken equal: $\Sigma_{\rm R}=\Sigma_{\rm L}=\Sigma$, and have the following energy-dependence \cite{Sasada}:

\begin{equation}\label{SelfEnergy}
\Sigma=\frac{E}{2}-i\sqrt{1-\left(\frac{E}{2}\right)^2}
\end{equation}

By ``central'' region, we mean the site $|0\rangle$ connected to the cavity and the leads at the same time. In order to have full transparency ($\mathcal{T}=1$) at all the Fermi energies, the energy $\epsilon_0$ should be equal to that of the lead sites: $\epsilon_0=0$. Note that we kept $\epsilon_0$ explicit in Eq.~\eqref{GF} despite its vanishing value, in order to better see the partition of the system and define the left and right self energies. The quantities $\Gamma_{\rm L}$ and $\Gamma_{\rm R}$ in \eqref{Landauer} characterize the coupling of the central region to the leads, and they are seen as linewidths. Their expressions are given by the imaginary part of the self energy of the leads:

\begin{equation}\label{couplingLR}
\Gamma_{{\rm L},{\rm R}}=-2\Im \Sigma_{{\rm L},{\rm R}}
\end{equation}

This partition into a central system and the surrounding leads and cavity is optional and not unique \cite{Darencet}. Our purpose is to obtain a scalar Green's function, which greatly simplifies the expression of the transmission function. Notice in the expression of the Green's function \eqref{GF}, that the off-channel cavity is taken into account through its effects on the central site, which can be understood from the decimation procedures \cite{Grosso,Lambert}; these effects may be contained into a self-energy:

\begin{equation}
\Sigma_{\rm d}=\frac{t_{\rm c}^2}{E-V_{\rm d}}
\end{equation}

Substitution of the above quantities into the Fisher-Lee formula \eqref{Landauer} yields the following simple expression for the transmission function of the system:

\begin{eqnarray}\label{Transmission}
\mathcal{T}(E)=\frac{1}{1+\left[t_{\rm c}^2/\Gamma(E-V_{\rm d})\right]^2}
\end{eqnarray}

\begin{figure}
\begin{center}
\includegraphics[scale=0.85]{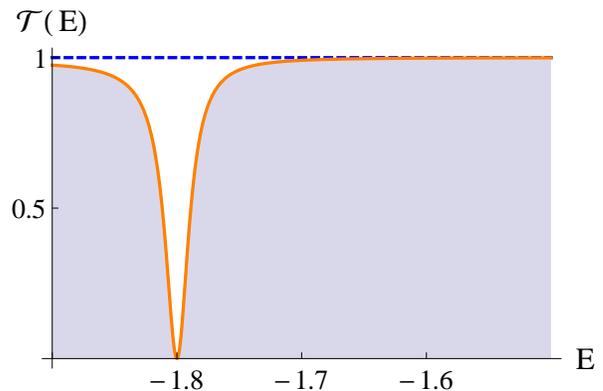}
\end{center}
\caption{(color online) Transmission profile in the presence of the off-channel cavity (orange curve) and without the cavity (dashed blue). The presence of the cavity changes drastically the transmission of the system and can even lead to cancellation when the Fermi energy equals  the cavity level (here $V_{\rm d}=-1.8$).}
\label{FigTransmission}
\end{figure}

The profile of the transmission is given in Fig. \ref{FigTransmission}. We notice that the presence of the cavity changes drastically the profile of the transmission and it even leads to its cancellation, which can be viewed as a  quantum destructive interference. The value at which  this cancellation appears is directly obtained from the transmission function, in Eq.~\eqref{Transmission} : as $E\rightarrow V_{\rm d}$, ${\mathcal T} \rightarrow 0$, and the width of the antiresonance is controlled by the coupling parameter $t_{\rm c}$. A first-order expansion near the energy $E=V_{\rm d}$ shows that the transmission function is parabolic:

\begin{equation}
\mathcal{T}(E)\sim \left(\frac{\Gamma}{t_{\rm c}^2}\right)^2 (E-V_{\rm d})^2
\end{equation}

\noindent and it becomes clear that the weaker the coupling to the cavity is ($t_{\rm c}^2\ll\Gamma $ ) the sharper the antiresonance is. To recover the full transmission of the system (as without the cavity) one may increase the potential $V_{\rm d}$ to infinity, since this implies that the cavity region becomes forbidden to the electrons moving across the waveguide. Decreasing the coupling $t_{\rm c}$ down to $0$ makes the antiresonance narrower but does not eliminate it; and setting $t_{\rm c}=0$ is a rather abrupt way to take the cavity off.

We just saw the effects of the coupling parameter $t_{\rm c}$ and the cavity level $V_{\rm d}$ on the transmission function, and clearly they have no influence on the symmetry of the profile. In order to induce some asymmetry we need to introduce an additional physical process, hence a new parameter. In what follows, we allow the electrons that flow through the waveguide to interact with phonons within two small regions located on both sides of the site coupled to the cavity.

\section{Effects of dephasing phonons}
\begin{figure*}
\includegraphics[scale=0.6]{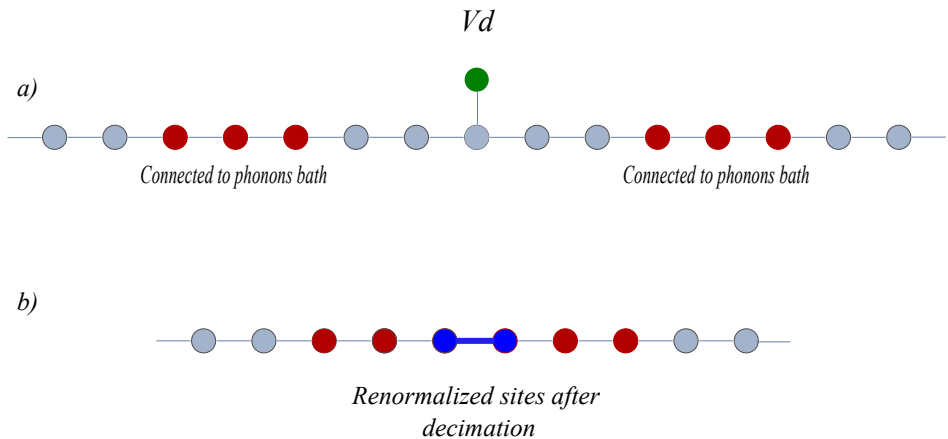}
\caption{(color online) Sketch of the decimation procedure $(a)\longrightarrow(b)$.}
\label{Graphdecimation}
\end{figure*}
In this section, to study the effects of electron-phonon interaction on the transport coefficients, we consider the presence of external phonon baths symmetrically located on both sides of the cavity and connected to $n=2\times N$ sites. The system remains essentially the same as that of the previous section where we considered phase-coherent transport, but we now assume that in some parts of the left and right leads, the electrons are interacting with phonons as shown in Fig. \ref{Graphdecimation}. The description of the new situation is done using a decimation procedure, which is described further below.

\subsection{Electron-phonon interaction}

Following Ref. \cite{Datta}, we make approximations to perform tractable calculations: we assume that the phonons, which form a bath of independent oscillators, are in equilibrium; we consider only one-phonon scattering processes (self-consistent Born approximation); the electron-phonon interaction is local and has no effect on the hopping coefficient \cite{Lake}. Further, we simplify our analysis by considering single-mode phonons, with energy $\hbar\omega_0$. The phonon-electron interaction Hamiltonian thus assumes a simple form and reads:

\begin{equation}
H_{\rm e-ph}=\lambda \sum_{x} |x\rangle \langle x| (b+b^\dagger)  
\end{equation}

\noindent where $b^\dagger$ and $b$ respectively are the second-quantized phonon creation and annihilation operators, and $\lambda$ is the electron-phonon interaction strength considered here as site-independent.

The approach to access the transmission function through the device in the presence of dephasing phonons starts with expressing the retarded (advanced) and lesser (greater) self energies as done in Refs.~\cite{Cresti,Cresti2,Cresti3,Caroli71,Caroli72}. These quantities are obtained in a closed form within the self-consistent Born approximation~\cite{Cresti,Haug}, which ensures current conservation \cite{Cresti,Wacker}.

\subsection{Decimation procedure}
To simplify the description of the system connected to the phonon baths on both sides of the central site, which is connected to the off-channel cavity, we use a decimation procedure \cite{Grosso,Lambert} as depicted in Figs.~\ref{Graphdecimation}$(a)$ and \ref{Graphdecimation}$(b)$: the red sites are connected to the phonon baths, and the green site is at the energy level of the off-channel cavity; the system thus is formed of two parts: the interacting region connected on its left and right sides to semi-infinite perfect 1D leads. The decimation consists in taking out the central part (five grey sites and the green one), and renormalizing the sites (and their links) to which it was attached. The renormalized sites are shown in blue. This procedure is exact in the sense that both systems $(a)$ and $(b)$ are equivalent for transmission since they share the same Green's function submatrix related to the remaining sites. The sites are renormalized as follows: we note $h$ the Hamiltonian of the part we take out and we define the Green's function $g=\frac{1}{E-h}$; then the Hamiltonian of the sites directly coupled to the removed part is:

\begin{equation}\label{DecimationEq}
h_{\rm r}=h_{\rm r}^{(0)}+\tau^\dagger \frac{1}{E-h}\tau
\end{equation}

\noindent where $h_{\rm r}$ is the new  Hamiltonian of the renormalized sites (a sub-Hamiltonian of the total one), $h_{\rm r}^{(0)}$ is the Hamiltonian of these sites in the absence of coupling to the central part, and $\tau$ is the coupling matrix between the central part and its surrounding sites ($\tau^\dagger$ is its complex conjugate). Equation (\ref{DecimationEq}) clearly shows that the Hamiltonian $h_{\rm r}$ is energy-dependent, and we recognize the form of a self-energy: $\tau^\dagger \frac{1}{E-h}\tau$. Actually the decimation procedure may be seen as the use of the concept of self-energy in the formalism of Landauer-Buttikker to take into account the effects of the reservoirs, or infinite leads (an application of such procedure may also be found in Refs.~\cite{Adel,Adel2}). Here, since the part of the system which is taken out is finite, the self-energy is real valued. The aim of the decimation procedure is to lower the dimension of the Green's matrix since it is numerically obtained using a recursive algorithm. Another interest of this procedure is that it simplifies the equations describing the phonon self-energy $\Sigma^{\rm ph}$, entering the definition of the retarded Green's function:

\begin{equation}\label{GreenFunction}
G=\frac{1}{E-H-\Sigma^{\rm leads}-\Sigma^{\rm ph}}
\end{equation}

\noindent where, after application of the decimation procedure, the Hamiltonian $H$ takes the form of an $n \times n$ matrix. The leads are characterized by their self-energy matrix $\Sigma^{\rm leads}$, which has only two nonzero elements: $\Sigma^{\rm leads}_{11} = \Sigma^{\rm leads}_{nn}=\Sigma$. The phonon self-energy is more complicated and it is obtained self-consistently with the Green's function. Within the Born approximation and the assumption that the effect of the phonons is local (their effect is restricted on single sites and does not change the hopping coefficients) the phonon self-energy reads:

\begin{equation}\label{SigmaPhonons}
\Sigma_{\rm ph}=\gamma^2 \mathcal{D}G  
\end{equation}

\noindent where $\mathcal{D}$ is a superoperator whose application to a matrix sets all its elements, except the diagonal ones, to zero, thus returning a diagonal matrix \cite{Cresti}. The parameter $\gamma$ characterizes the coupling between the phonon baths and the system, hence the dephasing process. Note that the Green's matrix \eqref{GreenFunction} and the phonon self-energy \eqref{SigmaPhonons} must be computed using a recursive process starting from an initial value of $\Sigma^{\rm ph}$, the computation being continued until convergence is reached to a good precision.

\subsection{Transmission function}

As explained in the works of Cresti and co-workers \cite{Cresti,Cresti2,Cresti3}, the transmission $\mathcal{T}$ at energy $E$ may be written as the sum of two contributions, which are called the coherent and incoherent terms:

\begin{eqnarray}
\mathcal{T}(E)=\mathcal{T}_{\rm coh}(E)+\mathcal{T}_{\rm inc}(E)  \label{TransiInc}
\end{eqnarray}

\noindent The coherent term originates in the leads' lesser self energies \cite{Cresti} and assumes the following form:

\begin{equation}\label{Tcoh}
\mathcal{T}_{\rm coh}(E)=\Gamma_{\rm L} Q_{1n}\Gamma_{\rm R}
\end{equation}

\noindent and the incoherent term, in the phonon's lesser self-energies \cite{Cresti}, assumes a similar form:

\begin{equation}\label{Tinc}
\mathcal{T}_{\rm inc}(E)=\Gamma_{\rm L} B_{1n} \Gamma_{\rm R} 
\end{equation}

\noindent In Eqs. \eqref{Tcoh} and \eqref{Tinc}, the couplings $\Gamma_{\rm L}$ and $\Gamma_{\rm R}$ are those defined in Eq.~\eqref{couplingLR}, and the matrix elements $Q_{1n}$ and $B_{1n}$ (in the notations of Cresti \cite{Cresti}) are defined as follows. The matrix elements of $Q$ are:

\begin{equation}\label{TransiInc2}
Q_{ij}=|G_{ij}|^2  
\end{equation}

\noindent and the matrix $B$ reads:

\begin{equation}\label{TransiInc3}
B=\gamma^2 \frac{Q^2}{{\mathds 1}-\gamma^2 Q} 
\end{equation}

\begin{figure*}
\begin{center}$
\begin{array}{ccc}
\includegraphics[width=2.in]{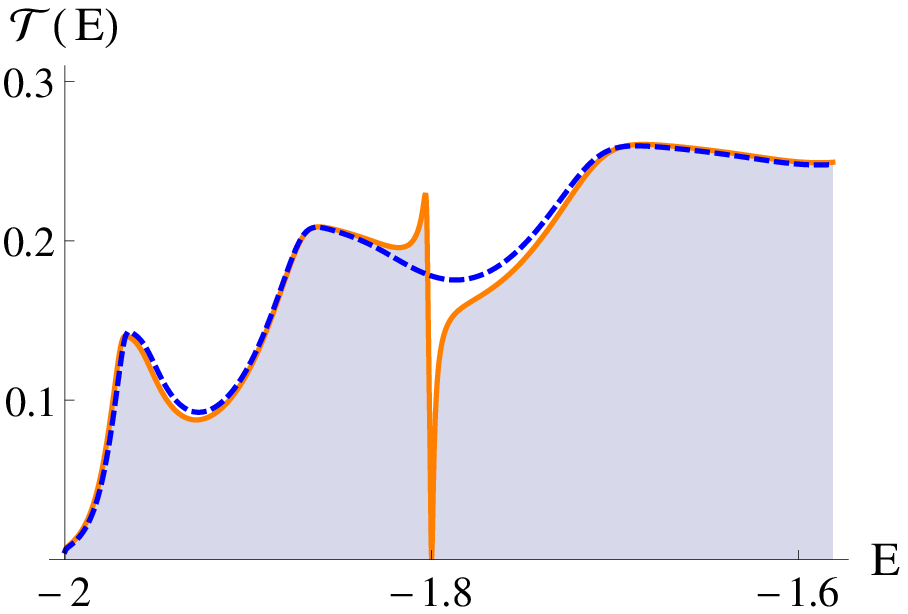} &
\includegraphics[width=2.in]{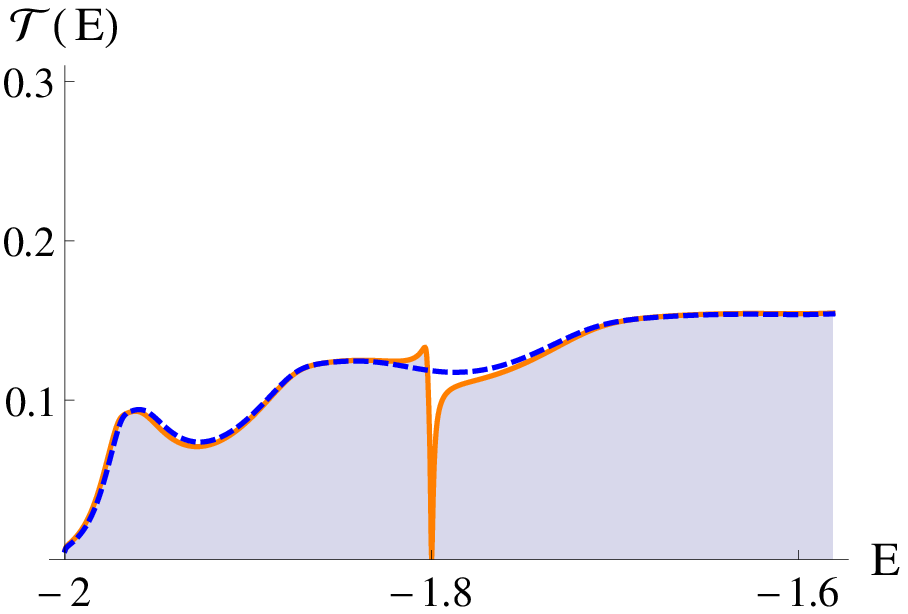} &
\includegraphics[width=2.in]{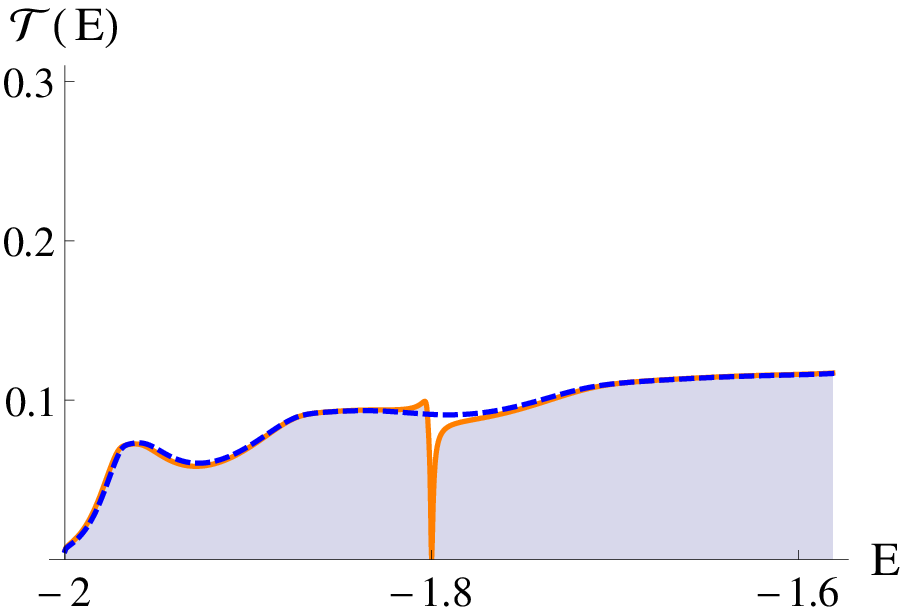}\\
\includegraphics[width=2.in]{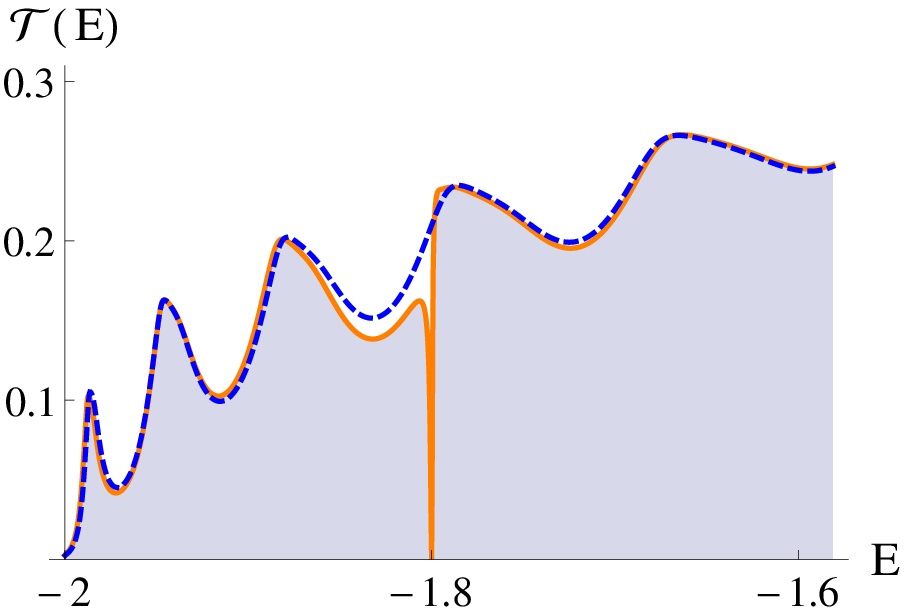} &
\includegraphics[width=2.in]{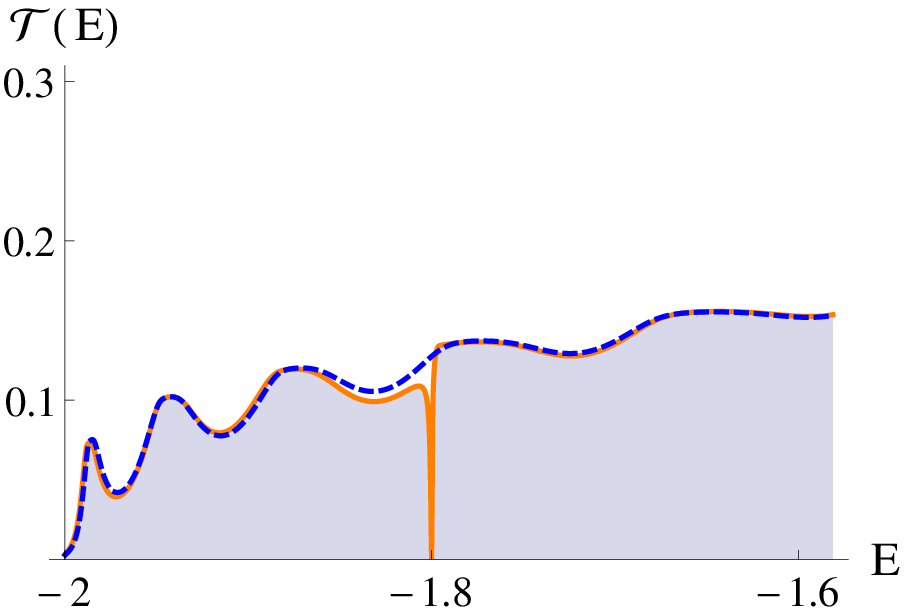} &
\includegraphics[width=2.in]{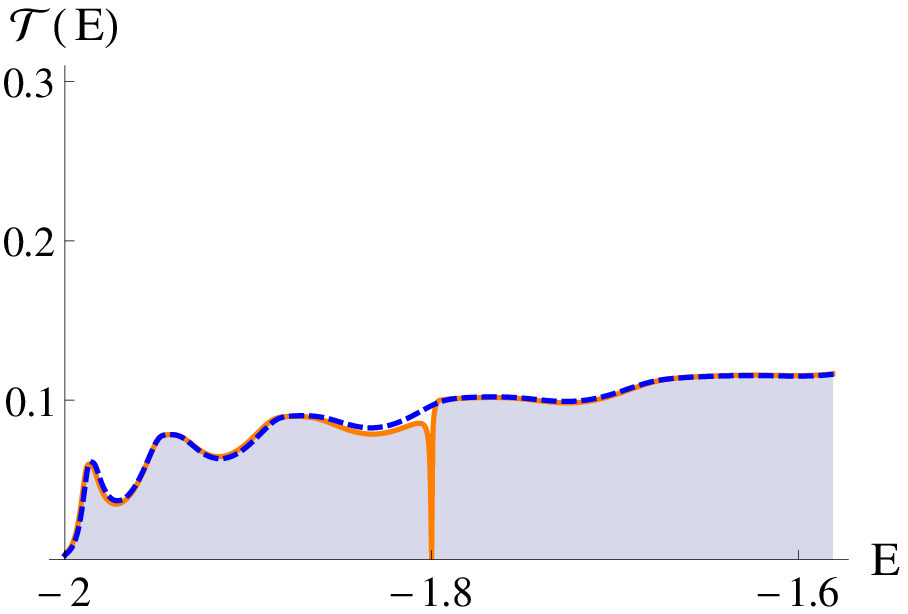}\\
\includegraphics[width=2.in]{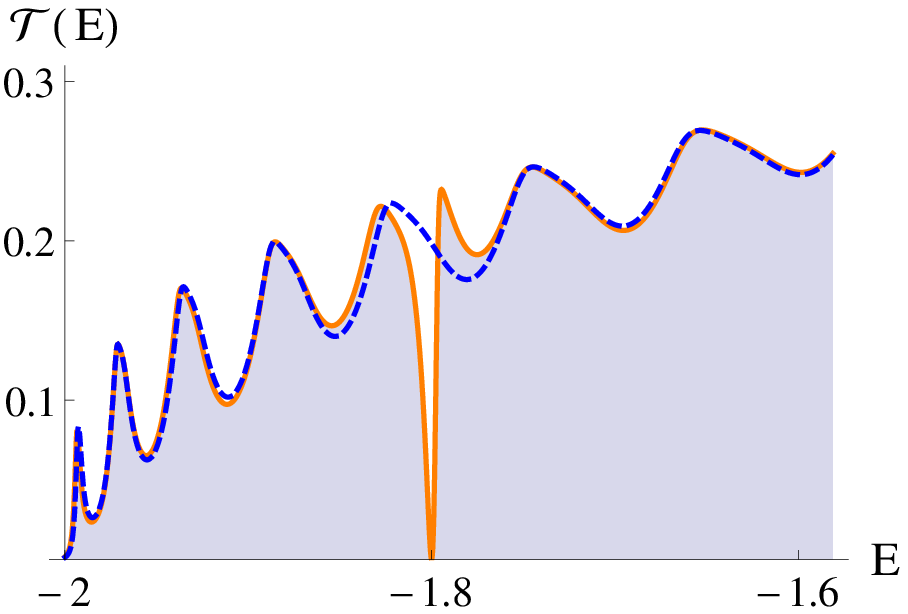} &
\includegraphics[width=2.in]{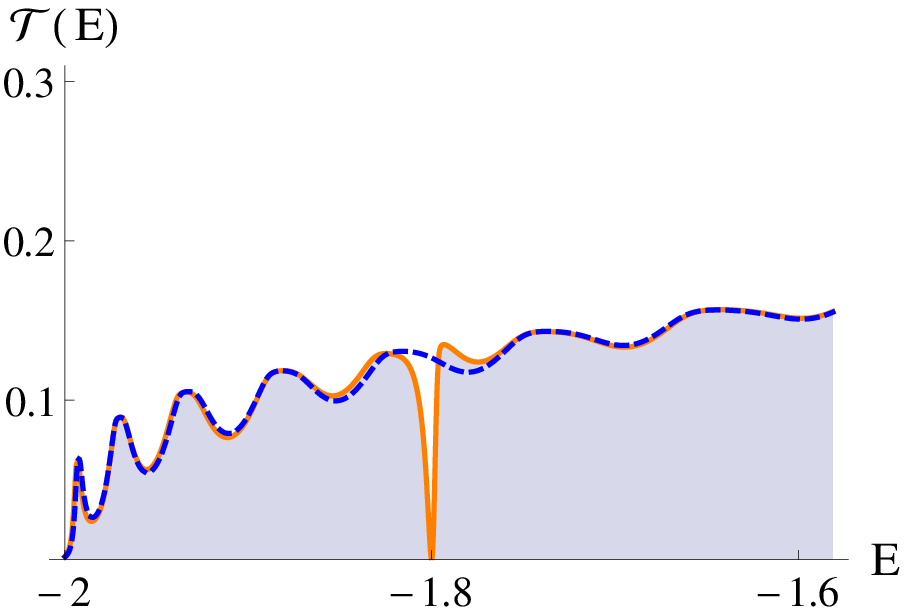} &
\includegraphics[width=2.in]{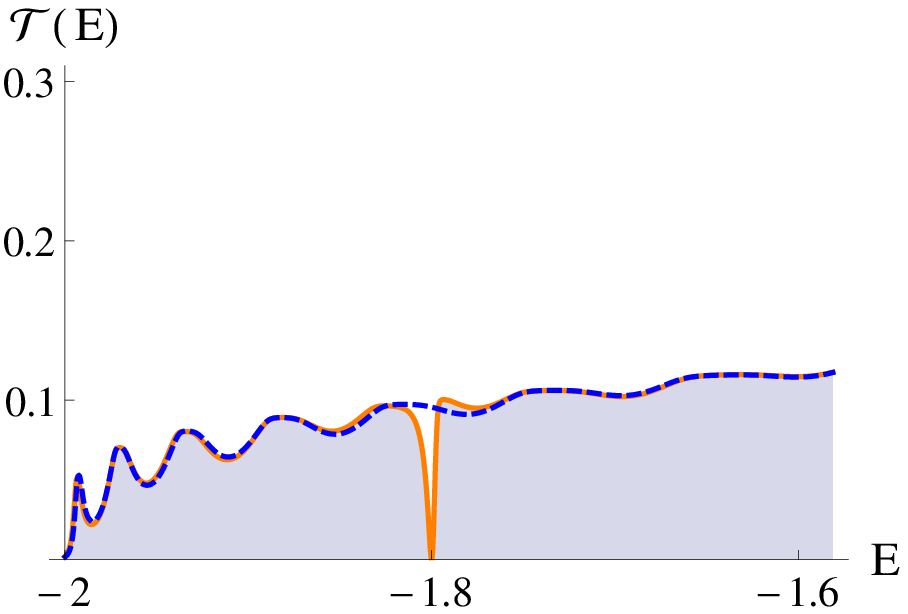}\\
\end{array}$
\end{center}
\caption{(color online) Transmission of the system for various lengths of the electron-phonon interacting regions and various distances between them. From left to right $l$ changes as $6a$, $10a$ and  $14a$, and from top to bottom the distance $L$ changes as $L=16a$, $26a$ $36a$, $a$ being the lattice step. For all the figures , $V_{\rm d}=-1.8$ and $\gamma=0.81$. The dashed curves represent the transmission of the system in the presence of phonons but without the off-channel cavity.}
\label{Dephasing}
\end{figure*}

We may now analyze the effects of the coupling of the electron waveguide to phonons on the transmission of electrons through the cavity. The numerical simulations based on an iterative procedure yield the results plotted in Fig.~\ref{Dephasing}. We discuss first the case without connection to the off-channel cavity. Comparison of Figs. \ref{FigTransmission} and \ref{Dephasing} shows that the presence of phonons drastically perturbs the transmission function: the amplitude is significantly reduced and oscillations appear. An increase of the size $l$ of the interacting regions reduces the transmission amplitude of the system, which reflects the dephasing effect; an increase of the distance $L$ between the two interacting regions induces more oscillations in the transmission profile: the system mimicks the behavior of a Fabry-Perot interferometer. Indeed, the presence of phonons renormalizes the potentials of the lattice sites with which they interact, via the real part of the self energy $\Sigma^{\rm ph}$, and since there are two distant electron-phonon interacting regions, the system shows oscillations of its transmission function. In the presence of an off-channel cavity, the transmission profile remains essentially the same, except near the cavity level where the transmission is canceled. Note that the transmission always cancels at the cavity level: the electron-phonon coupling, as described in our model, does not impact on the location of the antiresonance. The very interesting result we observe is the asymmetric aspect that the transmission profile acquires around the Fermi energy in the presence of dephasing phonons. The transmission of either high or low energy electrons can be reduced depending on whether the cancellation occurs at the local maximum or minimum of an oscillation: the system then acts as an energy-dependent carrier filter.

\begin{figure*}
\begin{center}$
\begin{array}{cccc}
\includegraphics[width=1.5in]{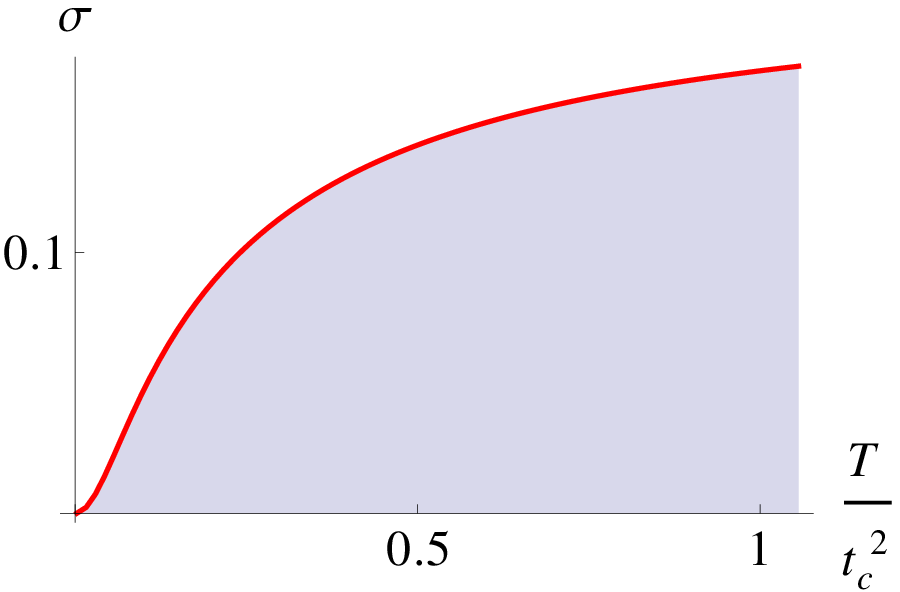} &
\includegraphics[width=1.5in]{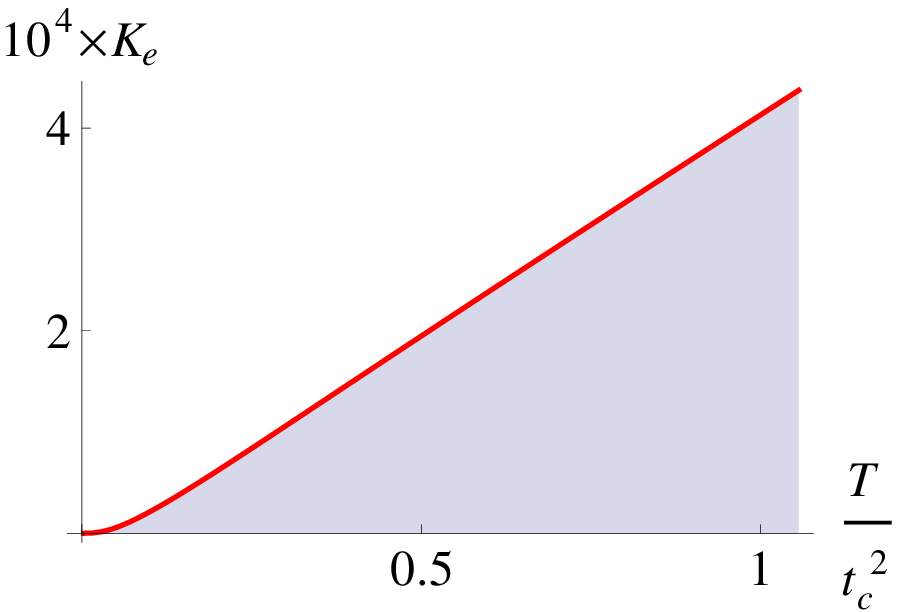} &
\includegraphics[width=1.5in]{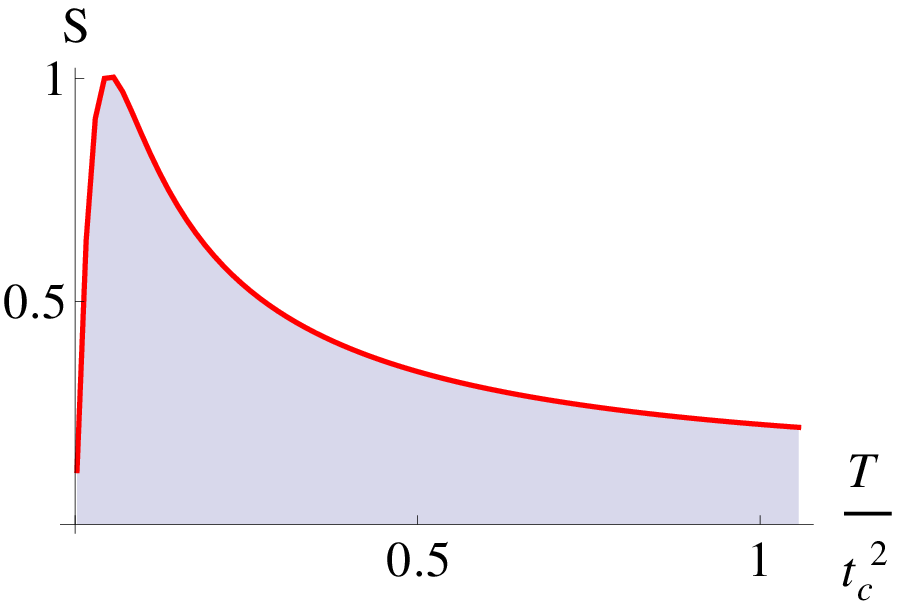}&
\includegraphics[width=1.5in]{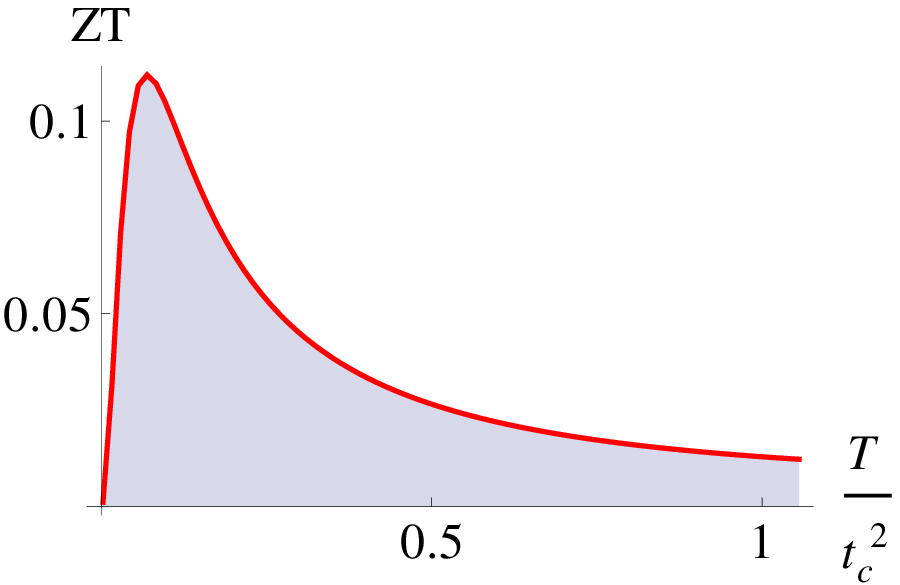} \\

\includegraphics[width=1.5in]{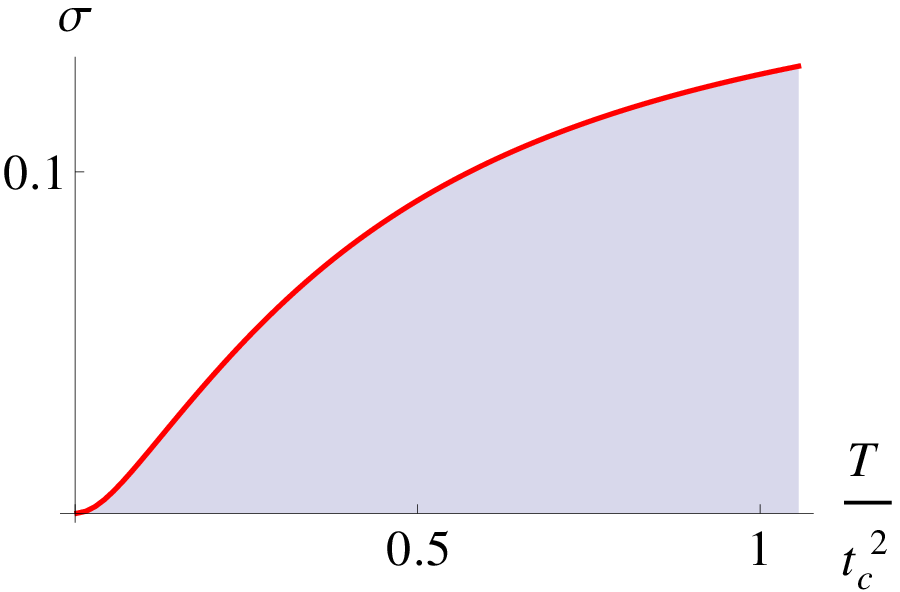} &
\includegraphics[width=1.5in]{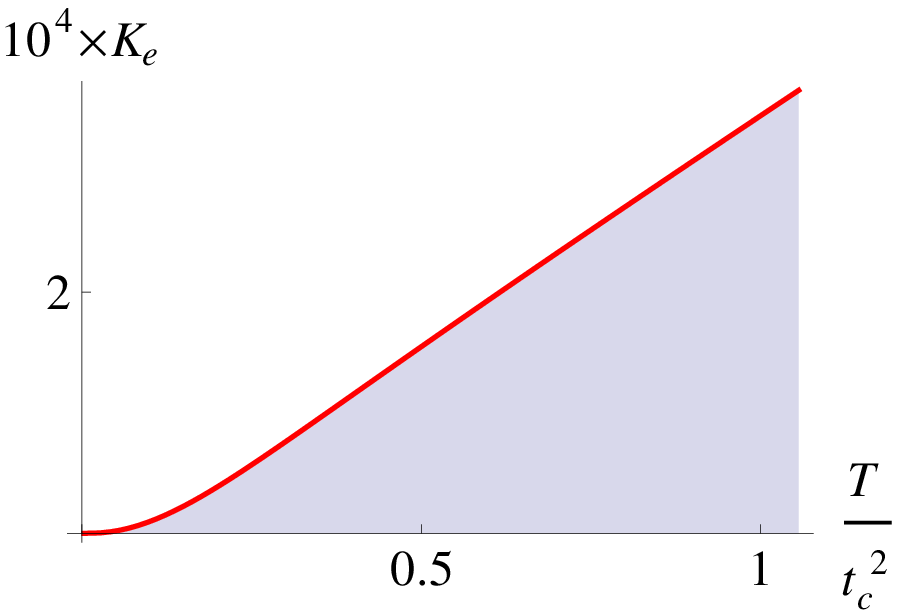} &
\includegraphics[width=1.5in]{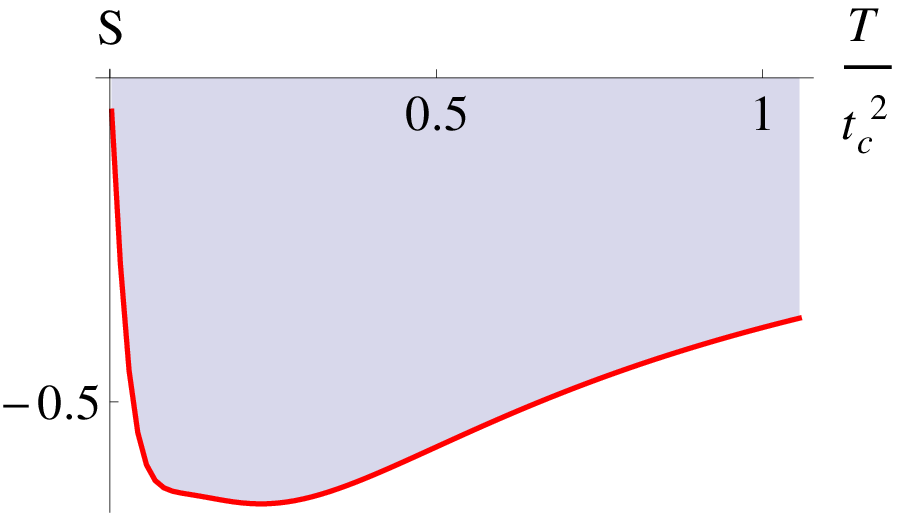}&
\includegraphics[width=1.5in]{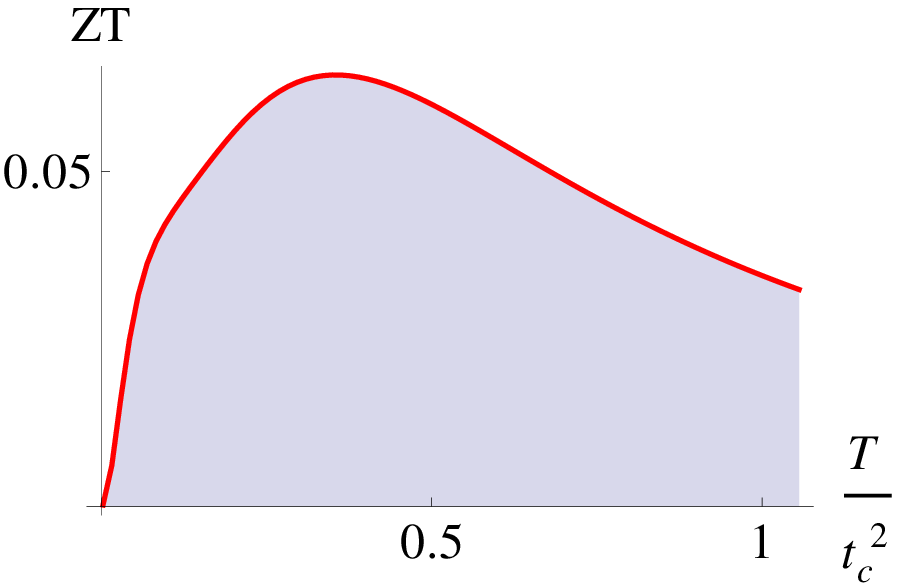} \\
\end{array}$
\end{center}
\caption{(color online) Temperature dependence of the transport coefficients for two different distances between the electron-phonon interacting regions. Top panels: $L=26 a$; bottom panels: $ L=16a$; and for both cases $l=6a$. The related transmission profiles are shown in Fig.\ref{Dephasing}}
\label{ThermoelectricFig}
\end{figure*}

\subsection{Thermoelectric coefficients in presence of phonons}

Once the profile of the transmission in known, the transport coefficients may be directly obtained by numerical integration of Eqs. \eqref{conductance}, \eqref{Seebeck}, and \eqref{Kappa}; these coefficients are shown as functions of temperature in Fig.~\ref{ThermoelectricFig}. We choose to analyze two different situations: one corresponds to a system more transparent to high-energy electrons, and the other is more transparent to low-energy electrons. In both cases we notice that the conductance $\sigma$ increases with temperature (note that in the linear regime, the temperature variation must remain small). This can be understood as follows: since the antiresonnance at the Fermi energy is extremely narrow, a temperature increase allows electrons with energy around the Fermi energy within a $k_{\rm B}T$ range, to be transmitted. The behavior of the electronic thermal conductivity $\kappa_{\rm e}$ is similar for the same reasons. 

Both profiles of thermopower $S$ as functions of temperature $T$ show an extremum. The difference of signs is due to the location of the antiresonance of the transmission function, which corresponds either to a minimum or a maximum of the oscillations around the Fermi energy: in the upper row of Fig.~\ref{ThermoelectricFig}, the system is more transparent to electrons with energy $E < E_{\rm F}$ whereas, in the lower row, it is more transparent to electrons with energy $E > E_{\rm F}$. This change of sign for the thermopower could also be anticipated from the formula \eqref{Seebeck}. Recall that this asymmetry is due to the modulation created by the two  electron-phonon interacting regions. Note that while the values of the figure of merit $ZT$ are modest in both cases, their shapes indicate that with appropriate adjustments of the location energy (which corresponds to the peak) and passband, very interesting developments are possible since the energy spectrum of the thermoelectric system is tunable. Indeed, this is a key point as regards the thermodynamic coupling of the system to its environment or other systems: minimization of the mismatch of the respective energy spectra of two systems permits a lowering of entropy production as these two systems get coupled (this is well known in the case of monochromatic photons interacting with a black body \cite{DeVos}).

\section{System connected to two off-channel cavities}
The analysis we did so far concerned various situations for which the electron-phonon interaction-induced asymmetry is at most typically of the order of the amplitude of the oscillations. We now see whether a stronger asymmetry is possible by simply connecting an additional off-channel cavity to the electron waveguide, still in the presence of dephasing phonons as in the previous case.

The two cavities are supposed to be small and each one contains only one energy level that is tunable with external gates. The presence of a second off-channel cavity adds a new parameter to the problem. When connected to the 1D system, the additional cavity also induces a cancellation of the transmission at an energy equal to that of its level. We now assume that the Fermi energy is equal to the energy level of one of the cavities and let the other level be free to be changed. The Hamiltonian of the new system is quite similar to that of the previous case: $H=H_{\ell} + H_{\rm c} +H_{\rm d} + H_{\rm e-ph}$, with

\begin{eqnarray}
H_{\ell} &=& \sum_{x=-\infty}^{+\infty}  -t(|x\rangle\langle x+1|+|x+1\rangle\langle x|)\\
H_{\rm d} &=& V_{\rm d} |d\rangle\langle d|+\epsilon_{\rm d} |d^\prime\rangle\langle d^\prime|\\
H_{\rm c} &=& t_{\rm c}|0\rangle\langle d|+t_{\rm c}|l\rangle\langle d^\prime|+ \mbox{h.c} 
\end{eqnarray} 

\noindent where $H_{\ell}$ is the Hamiltonian of the electron waveguide, $H_{\rm d}$ is the Hamiltonian of the two cavities with $V_{\rm d}$ and $\epsilon_{\rm d}$ being their levels, and $ H_{\rm c}$ is the Hamiltonian that characterizes the coupling between the waveguide and the two cavities. For simplicity, we choose the same coupling parameter $t_{\rm c}$ for both cavities. Here $l$ represents the index of the site to which the second level is connected (the first cavity is connected to the site with index $0$).

\begin{figure}
\begin{center}
 \includegraphics[scale=0.35]{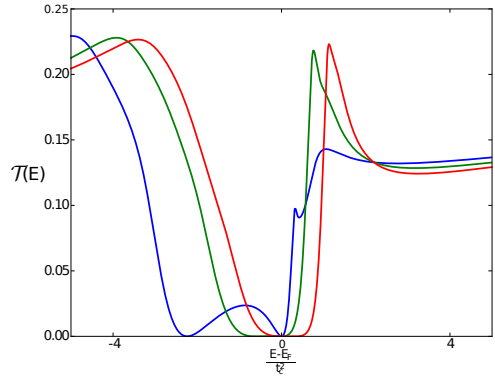}
\end{center}
\caption{(color online) Transmission profile in the presence of two cavities. The level of the first cavity is chosen to be $V_{\rm d} = E_{\rm F}=-1.8$. The energy level of the other cavity is $\epsilon_d$ such that: $(\epsilon_{\rm d}-E_{\rm F})/t_{\rm c}^2 = -2.24$ (blue curve); $(\epsilon_{\rm d}-E_{\rm F})/t_{\rm c}^2=-0.64$ (green curve); and $(\epsilon_{\rm d}-E_{\rm F})/t_{\rm c}^2 = 0.32$ (red curve); all with $t_{\rm c}=0.025$. The overlap of the two cancellations yields a large zero-transmission region. The green curve corresponds to the maximum of $ZT$ in Fig.\ref{ThermoelectricCoef}. The central noninteracting region contains $n=67$ sites and $\gamma=0.81$. The distance between the two cavities is $50 a$.} 
\label{ProfileOfTrans}
\end{figure}

The transmission function is obtained from Eqs. \eqref{TransiInc}, \eqref{TransiInc2}, and \eqref{TransiInc3}. By varying the level $\epsilon_{\rm d}$, we change the profile of the transmission ${\mathcal T}$ and obtain situations where the low or high  energy electrons in the vicinity of the Fermi level cannot be transmitted due to the overlap of the two antiresonances as can be seen in Fig. \ref{ProfileOfTrans}. The simulation is performed considering a system with a non-interacting central region of $n=67$ sites at a Fermi energy $E_{\rm F}=-1.8$ (equal to the cavity level $V_{\rm d}$). 

\begin{figure*}
\begin{center}$
\begin{array}{ccc}
\includegraphics[width=2.in]{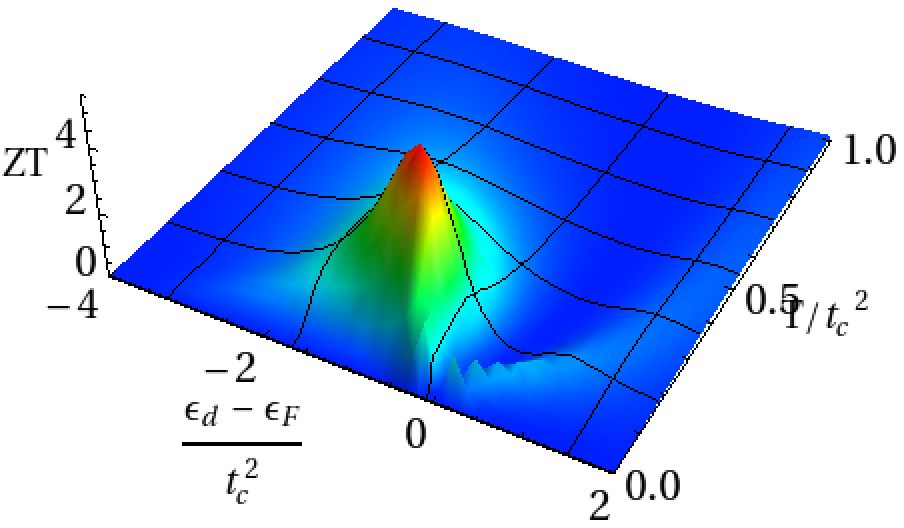}&
\includegraphics[width=2.in]{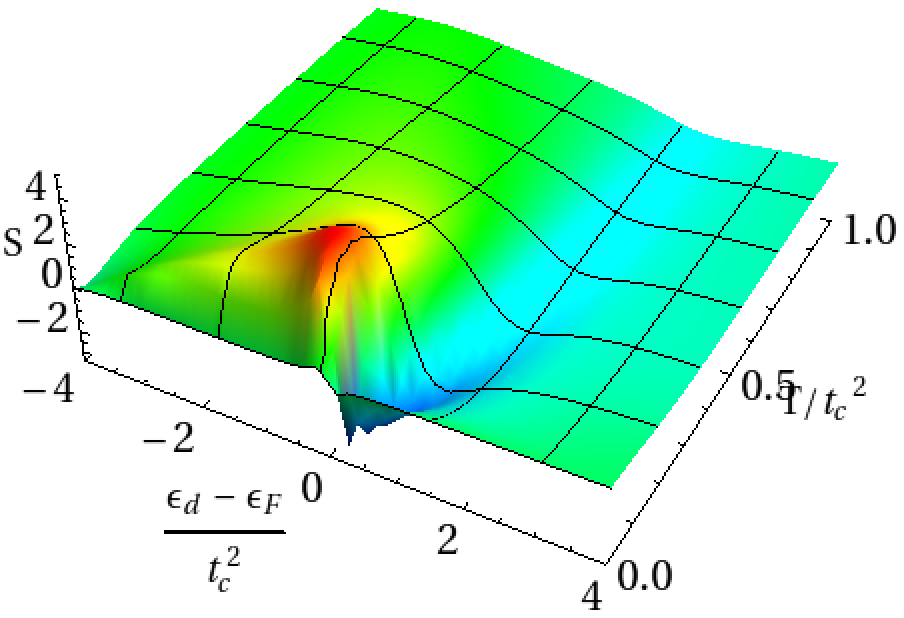}&
\includegraphics[width=2.in]{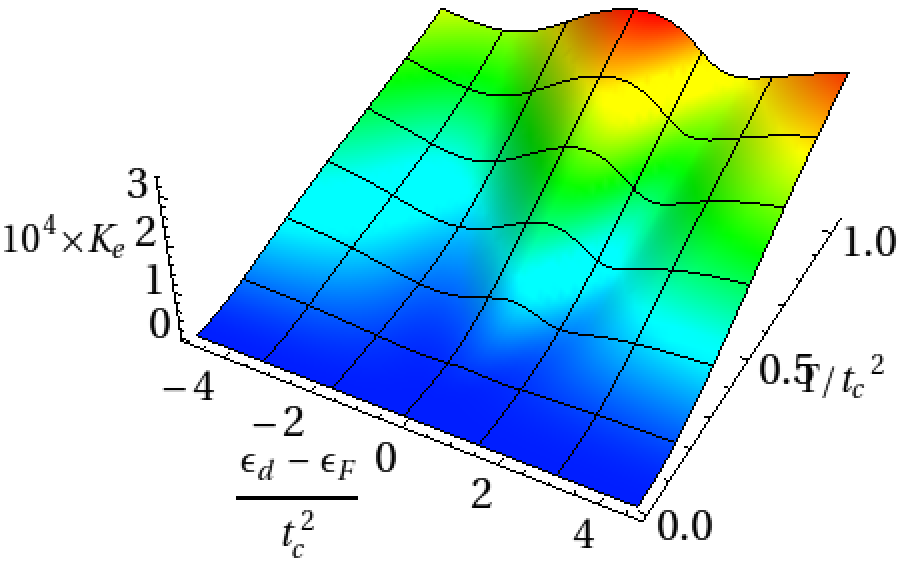}
\end{array}$
\end{center}
\caption{(color online) Transport coefficients as functions of the energy difference $\epsilon_{\rm d} - E_{\rm F}$, and temperature $T$.}
\label{ThermoelectricCoef}
\end{figure*}

In order to reach significant values of $ZT$, we investigate all the possible situations corresponding to various differences between the two cancellation energies, $\epsilon_{\rm d}-V_{\rm d}$, and we plot the figure of merit $ZT $ as a function of the temperature $T$. The numerical simulations are based on the same procedure as for the previous section. We only have to change the Hamiltonian of the central system by the new one including the additional cavity and iterate the algorithm in order to obtain in a self-consistent fashion the Green's matrix $G$ and the phonon self-energy $\Sigma^{\rm ph}$. The results are shown in Fig.~\ref{ThermoelectricCoef}. The figure of merit $ZT$ has a maximum at $ZT=5.38$ obtained for $(\epsilon_{\rm d}-E_{\rm F})/t_{\rm c}^2=-0.64$. This maximum and its location depend on the system's characteristics (the central region and the phonon's regions). We notice the presence of another local (but lower) maximum in the $ZT$ curve.

We also computed the thermopower $S$ as a function of the same parameters ($\epsilon_{\rm d}-E_{\rm F}$ and $T$). The Seebeck coefficient has a maximum which corresponds to the maximum of $ZT$ and a minimum which corresponds to the local maximum of $ZT$ factor. The sign of the Seebeck coefficient changes as $\epsilon_{\rm d}$ varies with respect to the Fermi energy: the thermopower is positive when low-energy electrons in the vicinity of the Fermi level are not transmitted (see Fig. \ref{ThermoelectricCoef}) whereas it becomes negative when high energy-electrons are not transmitted. Note that it is also possible that for the same configuration of the system, the Seebeck also changes sign because of temperature smearing which tends to give more weight either to lower- or higher-energy electrons as the temperature increases.

\section{Summary and concluding remarks}
Our study of the thermoelectric transport coefficients of a 1D electron waveguide connected to one and then two off-channel cavities, in the presence of dephasing phonons, is based on the Landauer-B\"uttiker formalism and the nonequilibrium Green's function techniques. We showed how to obtain a strongly asymmetric shape for the transmission function: first by coupling parts of the waveguide connected to an off-channel cavity to phonons, and then by coupling this system to an additional off-channel cavity. In our case, the transmission profile is characterized by oscillations, which trigger changes of sign of the thermopower as the location of the antiresonance of the transmission function corresponds either to a minimum or a maximum of these oscillations around the Fermi energy. This sign change shows that the system acts as an energy-selective filter: low-energy electrons in the vicinity of the Fermi level are not transmitted (the case of positive thermopower), and the higher-energy electrons are not transmitted (the case of negative thermopower).

As the transmission function vanishes we observe an enhancement of the thermoeletric coefficients. This result is in accordance with that of Ref.~\cite{Trocha}, where a strong enhancement of thermoelectric coefficients due to Coulomb correlations and destructive interference effects (hence vanishing of transmission) was found in a double quantum dot system. It is also interesting to see that while an enhancement of $ZT$ was found for transmission profiles that exhibit a sharp antiresonance, a significant enhancement is also found for a narrow transmission resonance \cite{Linke}, which is the opposite configuration. To some extent, this recalls Babinet's principle in optics.

Finally, as pointed out in the Introduction, from a thermodynamic viewpoint a thermoelectric system is a thermal engine to which a $ZT$-dependent energy conversion efficiency is associated. Recently some aspects of the question of efficiency at maximum power of low-dimensional thermoelectric systems have been discussed in terms of performance for quantum dots and 1D ballistic conductors \cite{Linke,Linke2,Grifoni}. These works pertain to the more general framework of finite-time thermodynamics \cite{Andresen1,Andresen2,Andresen3,Tu}, a field which contains a number of interesting open questions at the macroscopic level \cite{vandenBroeck1,Apertet2,Apertet3,vandenBroeck2}, which must also be addressed at the mesoscopic level; these questions are related to the location of irreversibility sources in the system as a whole and the impact on the efficiency at maximum power, two particular cases being those of endoreversible and exoreversible engines. The question of irreversibility at the mesoscopic level is certainly not trivial since this requires a careful characterization of the coupling of a system to its environment.

\begin{acknowledgments}
We are pleased to thank Y. Apertet for a careful reading of the manuscript and insightful comments. This work was supported by a grant from the R\'egion Basse Normandie. We also acknowledge partial funding from the LabEx EMC3.
\end{acknowledgments}
  
\appendix
\section*{Green's matrix derived from the decimation procedure}
As an example, we give the matrices defining the Green's function for the model presented in Fig.~\ref{Graphdecimation}. First we define the Hamiltonian $h$ of the part we want to take out:

\begin{equation}
h=\begin{bmatrix}  0 & -1 &   0   &  0    &  0 &  0   \\ 
                  -1 &  0 &  -1   &  0    &  0 &  0   \\
                   0 & -1 &   \epsilon_0   & -t_{\rm c}  & -1 &  0   \\
                   0 &  0 &  -t_{\rm c} &  V_{\rm d}  &  0 &  0   \\
                   0 &  0 &  -1   &  0    &  0 & -1   \\
                   0 &  0 &   0   &  0    & -1 &  0   \\
\end{bmatrix}
\end{equation}

\noindent where the diagonal elements $h_{ii}$ represent the potentials on the sites of the central part including the cavity level, and the off-diagonal elements $h_{ij}$ show if site $i$ is connected to site $j$ and with which strength.

To perform the decimation procedure, we take out this part and renormalize the sites to which it was attached. We first express the Green's function $g=\frac{1}{E-h}$. After decimation, we obtain the system shown in Fig.~\ref{Graphdecimation}b with the Hamiltonian of the region between the leads, which reads:

\begin{equation}
H=\begin{bmatrix}  0 & -1& 0 & 0 & 0 & 0 \\ 
                -1 & 0 &-1 & 0 & 0 & 0 \\
                 0 & -1& g_{11} & g_{1n}& 0 & 0 \\
                 0 & 0 & g_{n1}& g_{nn} &-1 & 0  \\
                 0 & 0 & 0 & -1& 0 &-1  \\
                 0 & 0 & 0 & 0 & -1& 0  
 \end{bmatrix}
\end{equation}

\noindent Note that we used the fact that the coupling of the central system to its environment (lead with phonons) is  $t=1$; otherwise $t^2g_{ij}$ must be substituted to $g_{ij}$.

The self energy matrix of the leads reads:

\begin{equation}
\Sigma^{\rm leads}=\begin{bmatrix}  \Sigma & 0 & 0 & 0 & 0 & 0 \\ 
                               0 & 0 & 0 & 0 & 0 & 0 \\
                               0 & 0 & 0 & 0 & 0 & 0 \\
                               0 & 0 & 0 & 0 & 0 & 0  \\
                               0 & 0 & 0 & 0 & 0 & 0 \\
                               0 & 0 & 0 & 0 & 0 & \Sigma  
 \end{bmatrix}
\end{equation}

\noindent where $\Sigma$ is the self-energy of each semi-infinite perfect lead given in Eq.~\eqref{SelfEnergy}. The recursion process that drives the numerical computation starts with the initial value of $\Sigma^{\rm ph}= \Sigma^{\rm leads}$ and continues 
with equations \eqref{SigmaPhonons} and ~\eqref{GreenFunction}.


\begin{thebibliography}{20}
\bibitem{remark1} It is important to note that $\kappa_{\rm e}$ is the electronic thermal conductivity under open-circuit condition, i.e. for zero electric current.
\bibitem{remark2} With a small cavity size, we assume that the other energy levels are far too high to be accessible.
\end{thebibliography}
\end{document}